# Voxel-Matching NORDIC: Non-local patch formation by time-series similarity increases tSNR in high-resolution BOLD fMRI

Alessandro Nigi[1]; Natalia Petridou[2,3], Jeroen C.W. Siero[1,2,3]

[1] *Department of Radiology, Center for Image Sciences, Advanced MRI Gradient Hardware and Methods Group, University Medical Center Utrecht, The Netherlands*

[2] *Department of Radiology, Center for Image Sciences, Translational Neuroimaging Group, University Medical Center Utrecht, The Netherlands*

[3] *Spinoza Center for Neuroimaging Amsterdam, The Netherlands*

## ABSTRACT

**Submillimeter functional magnetic resonance imaging (fMRI) based on blood-oxygenation-level-dependent (BOLD) signal enables the study of brain function at the submillimeter level, uncovering insights into fine-scale organisations like cortical layers and columns. However, its inherently low contrast-to-noise ratio (CNR) and signal-to-noise ratio (SNR) often limit its reliability and applicability. Noise Reduction with Distribution Corrected Principal Components Analysis (NORDIC PCA) is a locally low-rank denoising algorithm that reduces thermal noise levels in BOLD fMRI in a local patch manner. However, local patches often contain a mixture of signals from multiple tissues that negatively affect the low-rank structure of the patches, which limits the denoising capabilities of the algorithm. We propose an alternative approach for patch formation by gathering similar non-local voxels, dubbed voxel-matching (VM) NORDIC. The results on submillimeter-resolution BOLD fMRI data indicate that VM-NORDIC effectively promotes the low rank of the patches by boosting signal redundancy, allowing for more efficient noise attenuation. Moreover, the method barely affects spatial smoothness due to the non-local voxel selection based on time-series similarity. In particular, VM-NORDIC outperforms standard NORDIC with default local patching (Standard-NORDIC) in terms of temporal SNR (tSNR) (~9-90% larger than Standard-NORDIC; ~23-250% larger than the original) and spatial smoothness estimates (~20% of the smoothness induced by Standard-NORDIC). These improvements are fundamental to improving the validity and precision of fMRI studies at submillimeter resolutions.**

## INTRODUCTION

Functional magnetic resonance imaging (fMRI) (Bandettini et al., 1992; Kwong et al., 1992; Ogawa et al., 1992) based on blood oxygenation level-dependent (BOLD) contrast is indispensable for depicting brain activity and functional connectivity. The development of ultra-high field (UHF) MRI systems (>= 7T) has allowed pushing the spatial resolution to the sub-millimeter scale by increasing the strength of the detectable signal (Uğurbil, 2018; Viessmann & Polimeni, 2021). High-resolution fMRI is especially essential in studying fine-scale structures at the mesoscopic level, like cortical columns and layers (Dumoulin et al., 2018; Jorgenson et al., 2015; Sudlow et al., 2015; Uğurbil, 2018). Nevertheless, fMRI short acquisition times, the small voxel sizes enabled by UHF, and the inherently small BOLD responses (~0.5-3%) lead to a low contrast-to-noise ratio (CNR) and signal-to-noise ratio (SNR), limiting the applicability and reliability of submillimeter fMRI (Dowdle et al., 2023; Fernandes et al., 2023; Viessmann & Polimeni, 2021; Vizioli et al., 2021). These drawbacks become prominent at submillimeter resolution regimes, where the detectable MR signal is relatively weak, making noise effects more dominant (Dowdle et al., 2023; Triantafyllou et al., 2005, 2011; Vizioli et al., 2021; Wald & Polimeni, 2017).

The prevalent type of noise in high-resolution fMRI is thermal noise, an i.i.d. zero-mean Gaussian-distributed noise (e.g. white noise) arising from the random fluctuations in the electrical resistance of the detectors or the magnetic field strength due to the thermal energy of the atomic nuclei in the body (Edelstein et al., 1986; Hoult & Richards, 2011). Thermal noise is practically and theoretically different from physiological noise. The influence of physiological noise decreases with increasing image resolution and derives from physiological phenomena like the heartbeat and respiration, which occur periodically over time (Bianciardi, Fukunaga, et al., 2009; Bianciardi, van Gelderen, et al., 2009; Cox et al., 2017; Hu & Kim, 1994). As such, it falls within the category of structured, non-white noise (Caballero-Gaudes & Reynolds, 2017; Kay et al., 2013; Lund et al., 2006; Murphy et al., 2013) and is the target of dedicated techniques such as independent component analysis (ICA) and its applications (e.g. ICA-AROMA) (Pruim et al., 2015). On the other hand, the influence of thermal noise *increases* at higher resolutions, shorter TRs and lower magnetic fields (Triantafyllou et al., 2005, 2011). In particular, it becomes dominant over physiological noise at ~0.8 mm isotropic voxel size, but it can still affect the quality of lower-resolution scans(Triantafyllou et al., 2005, 2011). Also, the application of parallel imaging to accelerate fMRI data acquisition introduces a non-uniform spatial amplification of thermal noise according to the geometry of the receive coils (g-factor) (Pruessmann et al., 1999) . Overall, thermal noise lowers CNR, image quality and temporal SNR (tSNR) by increasing the signal variance (Vizioli et al., 2021). Specifically, in BOLD fMRI, functional activity is identified via subtle changes in the voxel signal. Therefore, the random signal fluctuations associated with thermal noise interfere with the detection of the underlying signals of interest related to brain activation, leading to false positives and false negatives (Vizioli et al., 2021). Thermal noise removal is therefore crucial in fMRI analysis to ensure clinically and research-wise high-quality data that lead to accurate, reliable results.

A way to overcome the penalties due to thermal noise at high resolutions is to increase the static magnetic field strength even further (Uğurbil, 2018; Wald & Polimeni, 2017). Nevertheless, hardware costs and high field-related artefacts strictly limit this strategy[4]. An alternative is to decrease the noise level with appropriate data processing techniques. Spatially blurring the dataset with a smoothing filter is one popular method to increase the SNR (Triantafyllou et al., 2006). Smoothing is fast and easily applicable, but it comes at the expense of lowering spatial specificity, which is undesirable if the goal is to depict the activity of fine-scale structures with high precision (Triantafyllou et al., 2006; Vizioli et al., 2021). Temporal averaging is another approach often applied to cancel out random signal fluctuations. However, it also degrades spatial and temporal precision.

More recent denoising techniques use Principal Component Analysis (PCA) to identify and nullify the data components indistinguishable from thermal noise by exploiting signal redundancy across volumes. The repetitive acquisitions inherent to diffusion MRI (dMRI) or fMRI provide the data with explicit signal redundancy. That is, the multiple volumes report the same underlying biological environment, allowing the dataset to have a locally low-rank (LLR) structure (Fernandes et al., 2023; Moeller et al., 2021; Veraart, Fieremans, et al., 2016). A locally low-rank structure means that in small patches of an image, just a few principal components contain most of the signal-related variance. This low rankness allows approximating the intensities of the voxels by a smaller number of signal principal components to preserve the details and structure of the original image while removing noise (Fernandes et al., 2023) .

In 2016, Veraart et al. introduced Marchenko-Pastur PCA (MPPCA) to denoise dMRI. MPPCA automatically estimates a threshold to suppress thermal noise principal components by exploiting the asymptotic properties of the eigen-spectrum of local data matrices corrupted by thermal noise. Such spectra follow the well-known Marchenko-Pastur (MP) distribution(Marčenko & Pastur, 1967) , whose bounds depend on the variance of the data and is used in MPPCA as the threshold to nullify the noise components (Veraart, Fieremans, et al., 2016; Veraart, Novikov, et al., 2016). MPPCA processes the datasets patch-wise, assuming the noise level to be constant within the patch. However, accelerated data do not always meet this assumption, potentially leading to an incorrect estimation of the noise variance (Pruessmann et al., 1999). Further, the bounds of the MP distribution are well defined only in the asymptotic limit (i.e. with a large amount of data), which is often an unrealistic condition with a finite amount of data (Dowdle et al., 2023; Veraart, Novikov, et al., 2016). The prevalent consequences of these drawbacks are an incorrect truncation of the singular values and the spatial smoothing of the data due to an excessive dimensionality reduction of the patches, which decreases the characteristic differences between the time series of adjacent voxels (Moeller et al., 2021). Smoothing is highly unwanted in sophisticated denoising methods such as PCA-based techniques, as it can lead to erroneous results and biased statistical analyses. Furthermore, as unwanted smoothing can seemingly increase SNR at the expense of spatial integrity, one could simply spatially smooth the data rather than applying advanced PCA-based denoising techniques.

Moeller et al. partly overcame the MPPCA limitations by introducing NOise Reduction with Distribution Corrected (NORDIC) PCA for dMRI (Moeller et al., 2021). The main difference between the two methods is that NORDIC estimates the noise variance either from a noise scan or the complex data via MPPCA and applies it to compute the noise threshold via Monte Carlo Simulations of the Gaussian noise in the data (Moeller et al., 2021; Vizioli et al., 2021). Moreover, NORDIC uses MPPCA to compute local noise levels to spatially flatten the noise due to the g-factor and ensure constant noise levels within any patch. The noise flattening is eventually reversed after denoising to re-establish the original signal spatial variation. The authors reported that NORDIC surpassed MPPCA in terms of the degree of noise reduction and preservation of detail on dMRI data (Moeller et al., 2021).

The NORDIC authors argued that fMRI can also benefit from LLR denoising thanks to the high redundancy of the data (Moeller et al., 2021; Vizioli et al., 2021). In this case, the method aims at denoising the time courses of the voxels. Several studies reported that NORDIC effectively removed the noise in the fMRI data while preserving the temporal structure of the underlying brain activity, leading to improved results in fMRI-based analyses, such as brain activation mapping and connectivity analysis (Knudsen et al., 2023; Moeller et al., 2021; Vizioli et al., 2021). Especially, Vizioli et al. showed that NORDIC outperforms MPPCA on fMRI data regarding tSNR levels, spatial smoothness and functional activation maps on data with a wide range of resolutions, scanning parameters and tasks (Vizioli et al., 2021).

However, local patches may represent a downside for NORDIC, as they often contain heterogeneous signals from multiple tissues that can contaminate signal redundancy, degrading the low-rank structure of the data (Zhao et al., 2022). Another disadvantage is that, as with MPPCA, NORDIC slightly smooths the data due to its LLR approach. Recently, Zhao et al. tackled these problems by developing a *non-local* low-rank (NLLR) PCA denoising method for dMRI. The proposed method uses matrices of 3D (2D spatial + 1D diffusion direction) similar non-local patches (Coupe et al., 2008; Dabov et al., 2007; Zhao et al., 2022). They argued that gathering similar patches into a single matrix better exploits signal redundancy and promotes the low-rankness of the dMRI data. The authors reported that their NLLR-based patching method led to a more effective dMRI denoising than MPPCA(Zhao et al., 2022).

As mentioned, NORDIC for fMRI also relies on signal redundancy across volumes to define principal components and thus could benefit from non-local patch formation (Moeller et al., 2021; Vizioli et al., 2021). In the present work, we determine the denoising performance of an alternative patching method for NORDIC that uses non-local similar voxels based on the time-series correlation to construct low-rank temporal patches (voxel matching, VM; dubbed VM-NORDIC). We hypothesize that the higher temporal homogeneity of the resulting patches compared to the default local patches (Standard NORDIC) leads to a more robust estimation of the low-rank structure of the dataset after PCA. Gathering voxels with similar time series boosts the degree of redundancy within the patch, amplifying the principal components related to the signal. Additionally, denoising together voxels from different locations in the dataset avoids spatial over-smoothing. As such, we expect VM-NORDIC to produce higher noise attenuation and greater preservation of spatial detail than Standard-NORDIC while applying the same parameter-free threshold on high-resolution fMRI data. VM-NORDIC retains NORDIC's parameter-free threshold while introducing data-driven patch-formation hyperparameters. We assess and compare the performance of VM-NORDIC with Standard-NORDIC on a group of submillimeter resting-state BOLD fMRI datasets at 7 tesla in terms of fundamental metrics for fMRI, including tSNR and global spatial smoothness.

## THEORY

*PCA and SVD*

Let $\mathbf{Y}$ be an $M \times Q$ complex-valued volumetric fMRI measurement Casorati matrix, with the rows representing $M$ voxels and the columns representing $Q$ MR signal samples for each voxel (e.g. time points of the BOLD data). We expect the fMRI data to have a low-rank representation because multiple volumes are acquired over time. That is, $Q$ is large enough to represent $\mathbf{Y}$ with a combination of a few linearly independent sources, or principal components $\mathbf{P} \ll \mathbf{R}$, with $\mathbf{R} = \mathrm{rank}(\mathbf{Y}) = \min\{M, Q\} \gg 1$ and $\mathbf{P} = \mathrm{rank}(\mathbf{Y}_L)$, where $\mathbf{Y}_L$ is the low-rank representation of $\mathbf{Y}$. NORDIC estimates the principal components via singular value decomposition (SVD) of the data $\mathbf{Y} = \mathbf{U} \cdot \mathbf{S} \cdot \mathbf{V}^H$, where $\mathbf{U}$ is a unitary matrix whose columns are the left singular vectors of $\mathbf{Y}$ and contain information about the spatial structure of the signal. $\mathbf{V}$ is the unitary matrix whose columns are the right singular vectors of $\mathbf{Y}$, representing the temporal structure of the signal. The diagonal elements $S_{1,1}, \ldots, S_{Q,Q}$ of the matrix $\mathbf{S}$ (size $M \times Q$) are the singular values that represent the contribution of each source of signal or noise. Here, $S^2(j) = \lambda(j)$, $j \in \{1, \ldots, Q\}$, is the $j^{\text{th}}$ eigenvalue of the temporal sample covariance matrix $\mathbf{\Sigma} = \frac{1}{M}\mathbf{Y}^H \cdot \mathbf{Y} = \frac{1}{M}\mathbf{V} \cdot \mathbf{S}^2 \cdot \mathbf{V}^H$, size $Q \times Q$, where $\mathbf{V}$ contains the temporal principal components.

For *noise-free* measurement data, $\mathbf{Y}$ can be effectively low-rank approximated by $\mathbf{P} \ll \mathbf{R}$ singular values, as the remaining $\mathbf{R} - \mathbf{P}$ singular values are zero (Moeller et al., 2021; Veraart, Novikov, et al., 2016). However, in a realistic case, $\mathbf{Y}$ is a *noisy* data matrix described by the model $\mathbf{Y} = \mathbf{X} + \mathbf{N}$, where $\mathbf{X}$ is the signal of interest and $\mathbf{N}$ is additive i.i.d. zero-mean Gaussian noise that propagates through *all* components, making all $\mathbf{R}$ eigenvalues non-zero, so that $\mathrm{rank}(\mathbf{Y}) = \mathbf{R}$.

Although the noise in each measurement is random, its effect on the spectrum of singular values becomes deterministic in the limit of $\mathbf{R} \gg \mathbf{P}$. In other words, if a noisy fMRI data matrix contains enough time points $Q$, the histogram of the noise-only components will follow the Marchenko-Pastur (MP) distribution, a well-understood asymptotic universal law from random matrix theory for random covariance matrices (Marčenko & Pastur, 1967). NORDIC PCA aims to estimate and eliminate the components within the MP distribution of a noisy dataset to attenuate thermal noise contributions. This result is achieved by numerically estimating the boundary between noise and signal singular values of a noisy data matrix and applying it as a threshold $\lambda_{\text{thr}}$ to nullify all the ordered singular values $\lambda(j) < \lambda_{\text{thr}}$ within the MP distribution (i.e. to cancel out all eigenvalues indistinguishable from zero-mean Gaussian-distributed noise). The truncated singular value matrix $\mathbf{S}_{\lambda\text{thr}}$ is then recombined with the original eigenvector matrices to reconstruct the low-rank denoised matrix $\mathbf{Y}_L = \mathbf{U} \cdot \mathbf{S}_{\lambda\text{thr}} \cdot \mathbf{V}^H$.

*Locally Low-Rank (LLR) model*

A locally low-rank model is a mathematical framework used to approximate a high-dimensional data set with a series of low-rank matrices (Lee et al., 2013) . This type of model allows for a more efficient and compact representation of the data and can be used for tasks such as dimensionality reduction, data compression, and denoising. The method of locally low-rank approximation typically involves breaking the matrix (here, the fMRI dataset) into smaller submatrices (local patches) and applying SVD to each submatrix (Lee et al., 2013). The locally low-rank property allows approximating each local region in the data by a low-dimensional subspace of fewer principal components. Of these components, some are dominated by the signal and others by noise - this separability is the basic principle of PCA-based denoising. Standard-NORDIC exploits the LLR properties of fMRI data to reduce the noise level via hard thresholding the ordered singular values of local patches (Lee et al., 2013).

*Degree of noise removal*

In VM-NORDIC, the threshold estimation and SVT are the same as for Standard-NORDIC. Nevertheless, the same threshold can be more effective at attenuating noise from a highly homogeneous patch rather than an inhomogeneous one. A higher homogeneity leads to a more defined low-rank structure of the data by making signal components more defined, which together allow approximating the data with a lower number of principal components. Better-defined principal components also diminish the risk of accidentally cancelling signal components and leaving noise components intact. However, as for Standard-NORDIC, noise propagates through all components of the spectrum of singular values (Knudsen et al., 2023; Vizioli et al., 2021) . Therefore, VM-NORDIC may lead to a higher noise attenuation than Standard-NORDIC but does not completely remove it from the data.

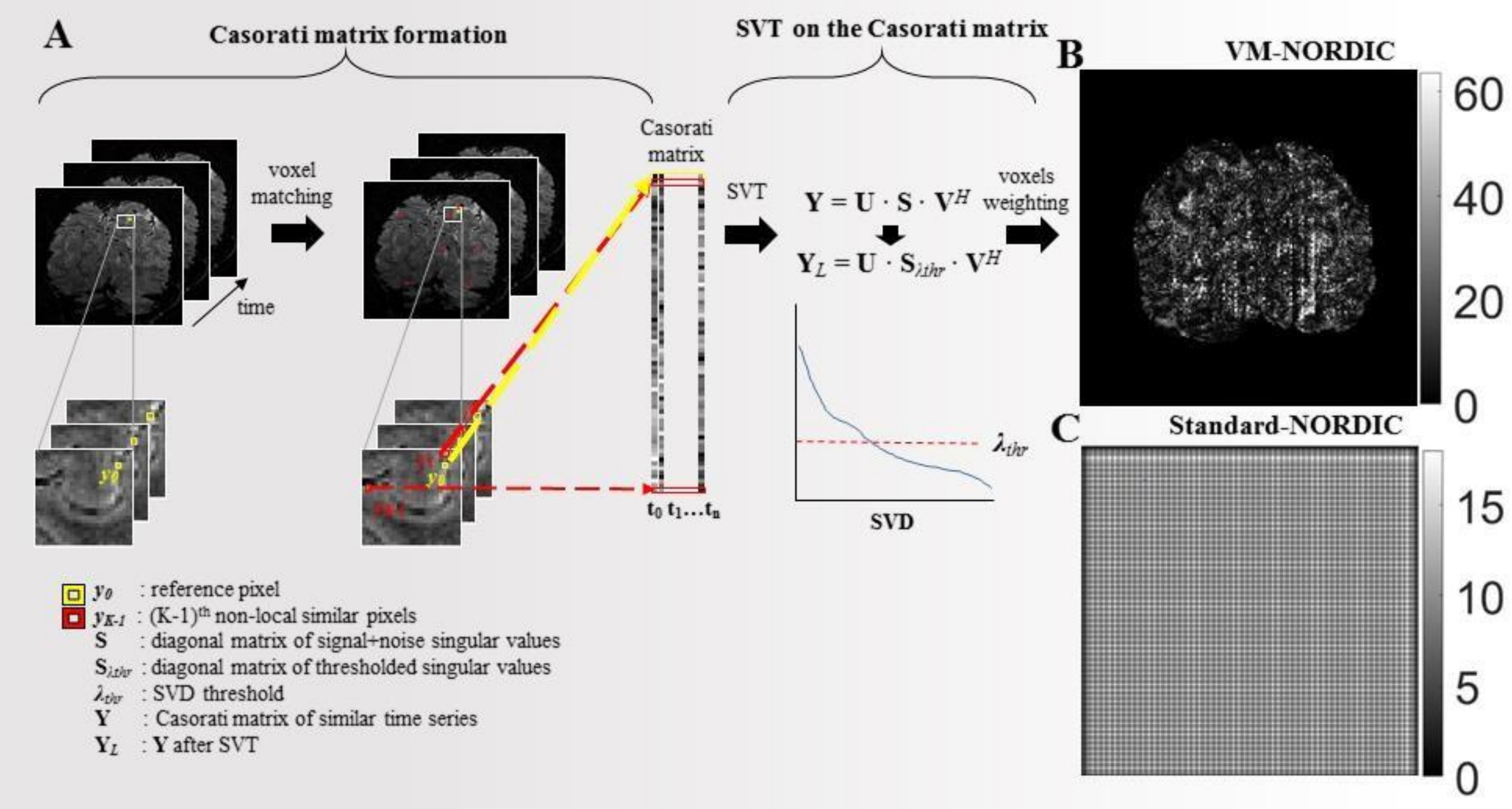

**Figure 1. A)** Flowchart of VM-NORDIC. For each chunk, $N$ reference voxels $\boldsymbol{y_0}$ are compared to all the voxels $\boldsymbol{y_m}$ in the chunk. For each reference voxel, the $K$-1 voxels with the highest similarity scores are grouped with the reference voxel into a $K{\times}Q$ Casorati matrix $\mathbf{Y}$. The matrix undergoes SVT to produce a denoised matrix $\mathbf{Y}_L$. Panels **B** and **C** show the weighting matrices and indicate the amount of time each voxel has been denoised for VM-NORDIC and Standard-NORDIC. The two panels also highlight the main differences in the patching approach between the two methods.

## METHODS AND MATERIALS

In the following sections, we first explain the implementation details of VM-NORDIC and highlight its differences from Standard-NORDIC. Then, we move to the methods we applied to acquire the data and analyse and compare the results from the two methods.

*Standard-NORDIC LLR model*

Standard-NORDIC relies on the LLR properties of fMRI data to reduce the noise level via hard thresholding the ordered singular values of local and spatially overlapping temporal patches. Neighboring voxels within a fixed $k_1 \times k_2 \times k_3$, usually with $k_1 = k_2 = k_3$, sliding-window are vectorized as $\mathbf{y}_t$ and gathered together to form a set of complex $M \times Q$ Casorati matrices of the form $\mathbf{Y} = [\mathbf{y}_1, \ldots \mathbf{y}_t, \ldots \mathbf{y}_Q]$, where $Q$ is the length of the time series and $M = k_1 \times k_2 \times k_3 \approx Q \cdot 11$, as reported by Veraart et al (see Figure 2A for an exemplar local patch) (Veraart, Novikov, et al., 2016; Vizioli et al., 2021). Each Casorati matrix undergoes SVT to truncate its singular value spectrum according to a parameter-free threshold $\lambda_{\text{thr}}$ (Moeller et al., 2021; Vizioli et al., 2021) . The result is a low-rank representation of the original Casorati matrix with a lower noise level. Finally, each matrix is reshaped into a 4D patch and relocated to its original spatial location.

*Proposed Voxel-Matching NORDIC*

In VM-NORDIC, patch formation occurs via grouping similar time series of non-local voxels. First, the dataset is divided into chunks of $X$ slices. Chunking the dataset is necessary to speed up the denoising process since processing all voxels at once may be too computationally heavy for most machines. For each chunk, VM-NORDIC computes the Euclidean distance between the time series of $N$ reference voxels and all the brain voxels in the chunk to assess their level of similarity. Alternatively, a correlation distance can be used to match according to shared temporal dynamics. For each reference voxel, the $K$-1 most similar voxels are vectorised and grouped with their reference voxel to form $N$ Casorati matrices of $K \times Q$ elements (Figure 1A), where $K$ is the (optimal) patch size as explained in the next section (see Figure 2B for an exemplar non-local patch). Each of the $N$ Casorati matrices undergoes SVT and is then transformed back into image space. Matching and SVT can run on both complex and magnitude data.

*Patch size optimization*

In VM-NORDIC, the patch size represents the number of vectorized time series per patch. Optimizing the patch size is a fundamental step in VM-NORDIC. Large patches may boost redundancy by gathering a higher number of similar time series. However, it also implies selecting time series that are less similar to the reference one, degrading the low rankness of the patch. On the other hand, small patches may be more homogeneous but not exhibit enough redundancy for reliable denoising. Therefore, in VM-NORDIC, the optimal patch size is a data-driven trade-off between the level of redundancy and the degree of similarity between time series. VM-NORDIC first makes an overestimated initial guess by multiplying the length of the time series by a factor $F > 1$. This step follows from the observation that the optimal patch size increases with increasing time points in the dataset (see Suppl. Figure 1 and 3). Then, the algorithm fine-tunes the initial guess via multiple dummy denoising runs to find the patch size $K$ that produces the highest mean tSNR score.

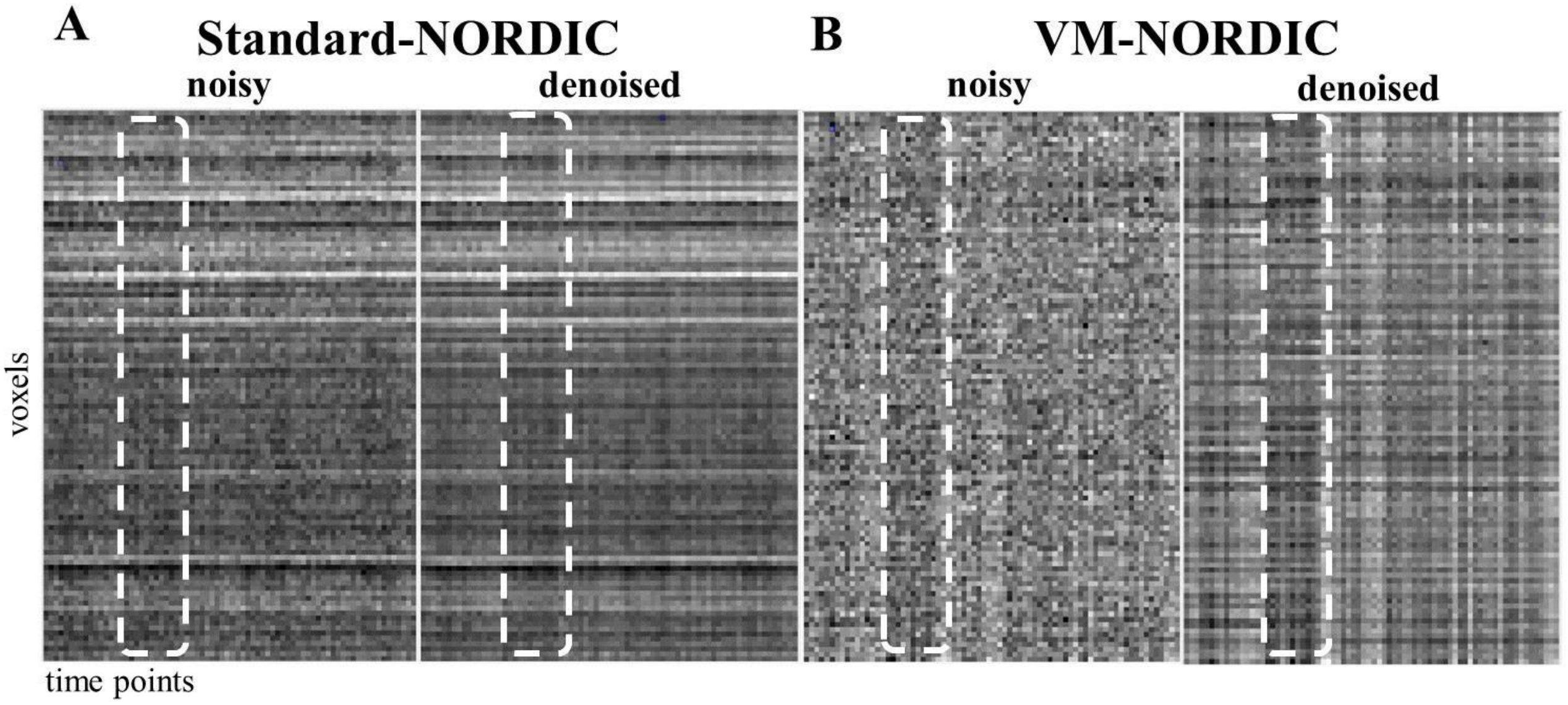

**Figure 2.** Exemplar Standard-NORDIC **(A)** and VM-NORDIC **(B)** patches before and after denoising. The dashed boxes indicate one of the temporal signal patterns shared by the timeseries (visible as vertical fading stripe). The VM-NORDIC patch is more homogenous than the Standard-NORDIC patch, allowing the denoising to better reveal and highlight the underlying common signal patterns.

*Threshold estimation*

The method for estimating the threshold $\lambda_{thr}$ is analogous in Standard-NORDIC and VM-NORDIC and is based on Monte Carlo simulations of noise matrices (Moeller et al., 2021; Vizioli et al., 2021). In VM-NORDIC, once the optimal patch size $K$ is selected, thermal noise is numerically simulated by generating multiple $K \times Q$ matrices filled with complex zero-mean Gaussian-distributed entries and with the same variance as the noise of the data. Then, each random matrix undergoes SVD, and the average highest singular value per decomposition is used as the threshold to separate noise from signal components in the data patches.

*Voxel averaging*

The proposed patching method implies that if a voxel is highly similar to $L$ reference voxels, it will undergo the denoising cycle $L$ times. The final voxel value is the weighted sum of its $L$ values generated after each denoising cycle it underwent. Figure 1B shows a weighting matrix for a representative slice. The values in the matrix represent the number of times the voxels are visited for denoising. This averaging procedure also functions as an additional denoising step by reducing the residual contributions of noise. A similar phenomenon occurs with Standard-NORDIC due to the overlapping patches (Figure 1C). The resulting 2D weighting matrices for the two methods highlight the core difference in the way they handle voxels.

*Brain masking*

VM-NORDIC applies a binary brain mask to the dataset to exclude the background from the denoising process using the FSL "bet" function on the time-averaged datasets (Popescu et al., 2012). Masking reduces the computational time by processing fewer voxels and ensures that the algorithm does not compute similarity scores between brain and background voxels, which could bias the SVD of the patch.

*Data chunks*

VM NORDIC divides the dataset into multiple chunks of 2D slices to further decrease the computation burden. The number of slices per chunk depends on a hardcoded value (~$10^5$) standing for the approximate number of voxels allowed per chunk. We chose this value based on the performance of the machines used for testing the algorithm and it can be adjusted for different needs. With these procedures, VM-NORDIC takes ~5-10 minutes to denoise a submillimeter BOLD fMRI dataset on our machines.

*Participants*

We acquired six resting-state datasets on three (two females and one male) healthy right-handed subjects (age range: 22-25) (see the "MRI imaging acquisition and processing" sections). All subjects had normal or corrected vision. All procedures were approved by the Medical Research Ethics Committee (METC) of UMC Utrecht in accordance with the Declaration of Helsinki; all participants gave written informed consent.

*MRI acquisition and processing*

All fMRI acquisitions were performed at 7T (Philips) using a 2×16-channel surface coil and covered the occipital lobe. A segmented 3D GE BOLD-EPI sequence was used. Participants were instructed to stay still, minimize movements and close their eyes. For the first dataset (A), the acquisition covered 30 coronal slices with TR/TE = 50/25ms, flip angle = 18°, segments = 1, SENSE factor (right-left, anterior-posterior) = 3/1, in-plane voxel size = 0.71mm$^2$, slice thickness = 0.8mm, volumes = 75, matrix size = 240×240, scan time = 4 min. For the second dataset (B), the acquisition covered 30 coronal slices with TR/TE = 65/31ms, flip angle = 18°, segments = 1, SENSE factor (right-left, anterior-posterior) = 4/1, in-plane voxel size = 0.71mm$^2$, slice thickness = 0.8mm, volumes = 75, matrix size = 240×240, scan time = 5.3 min. For the third dataset (C), the acquisition covered 44 coronal slices with TR/TE = 81/28ms, flip angle = 23°, segments = 1, SENSE factor (right-left, anterior-posterior) = 3/1, voxel size = 0.55mm$^3$ isotropic, volumes = 115, matrix size = 320×320, scan time = 15 min. For the fourth dataset (D), the acquisition covered 30 coronal slices with TR/TE = 96/31ms, flip angle = 20°, segments = 1, SENSE factor (right-left, anterior-posterior) = 3/1, in-plane voxel size = 0.5mm, slice thickness = 0.8mm, volumes = 75, matrix size = 352×352, scan time = 8 min. For the fifth dataset (E), the acquisition covered 33 coronal slices with TR/TE = 51/25ms, flip angle = 18°, segments = 1, SENSE factor (right-left, anterior-posterior) = 4/1, voxel size = 0.71mm$^3$ isotropic, volumes = 160, matrix size = 240×240, scan time = 10 min. For the sixth dataset (F), the acquisition covered 24 coronal slices with TR/TE = 98/28ms, flip angle = 23°, segments = 1, SENSE factor (right-left, anterior-posterior) = 3/1, voxel size = 0.45mm$^3$ isotropic, volumes = 100, matrix size = 320×320, scan time = 15 min.

*Experimental design*

This study consists of running VM-NORDIC and Standard-NORDIC on datasets with different resolutions and time points to assess which of the two methods leads to an overall more effective denoising in terms of key metrics like tSNR and spatial blurring. To ensure that the new patch size used in VM-NORDIC is not the major cause of differences in results between the two methods, we also performed multiple denoising runs with VM-NORDIC and Standard-NORDIC with different patch sizes. We then plotted Smoothness and tSNR scores Vs patch size to assess whether the two methods could produce similar results by only modifying the patch sizes.

*Metrics to assess denoising performance*

We assessed the denoising performance by comparing tSNR and spatial smoothness estimates between Standard-NORDIC and VM-NORDIC. The tSNR comes from dividing the mean temporal signal by the temporal standard deviation voxel-wise. Smoothness was estimated through the degree of spatiotemporal autocorrelation (FWHM) using the 3dFWHMx function from AFNI with the '-ACF', 'detrend' and 'automask' commands (Cox et al., 2017) . This function gives back a value in mm, representing the voxel spread (blurring). The spatial autocorrelation was estimated using a Gaussian + monoexponential decay mixed-model to account for possible long-tail autocorrelations (Cox et al., 2017; Vizioli et al., 2021).

## RESULTS

*Local vs. non-local patches*

Figure 2 displays two representative patches for Standard-NORDIC and VM-NORDIC before and after denoising. The Standard-NORDIC patch was cropped vertically in the figure to have the same size as the VM-NORDIC patch for easier comparison. The noisy Standard-NORDIC patch in Figure 2A includes similar voxels as well as voxels with different intensities and time courses. On the other hand, the non-local patch from VM-NORDIC in Figure 2B appears highly homogeneous and has clear common temporal patterns between time series. The denoised VM-NORDIC patch shows enhanced common temporal patterns compared to the Standard-NORDIC denoised patch.

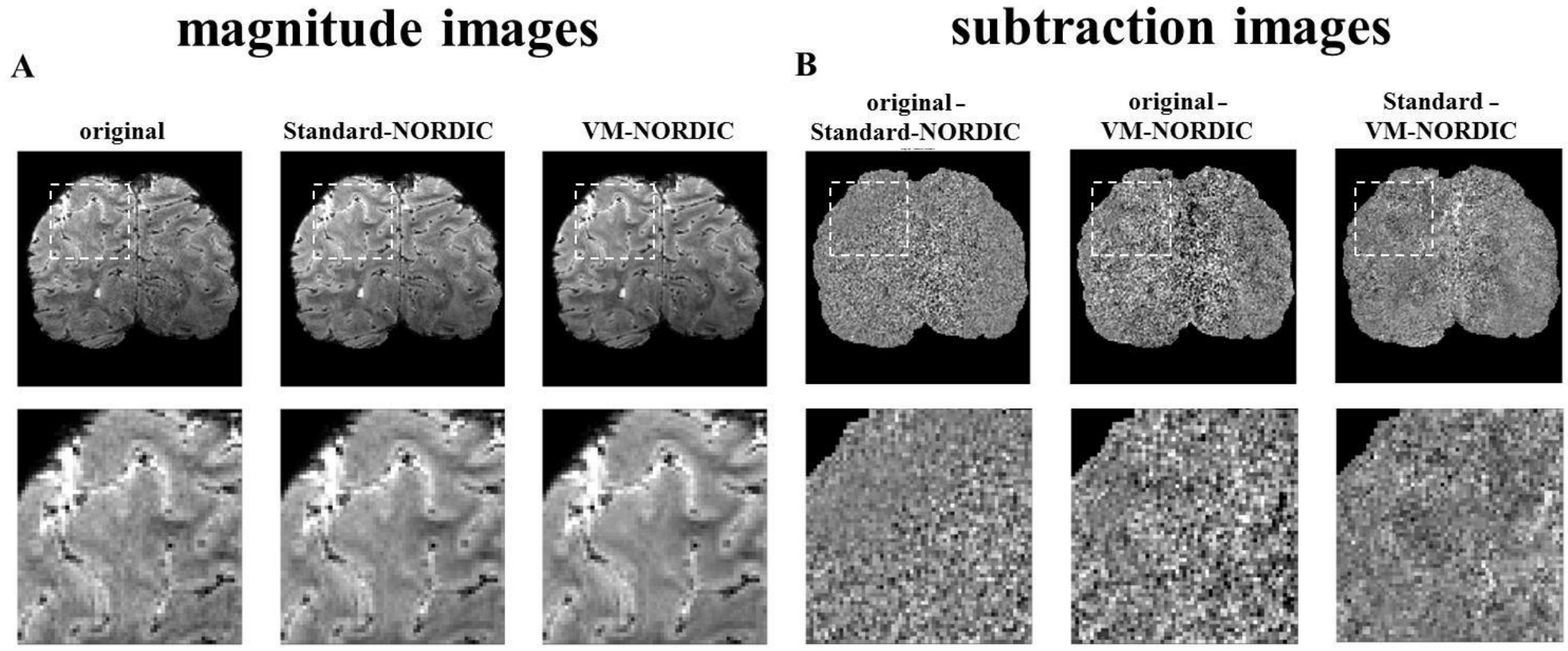

**Figure 3**. **A)** Magnitude images and their zoomed views for a representative slice of dataset B before and after denoising with Standard-NORDIC and VM-NORDIC. VM-NORDIC does not interfere with structural information and leads to more noticeable visual improvements. **B)** Subtraction images and their zoomed views of Standard-NORDIC versus the original data; VM-NORDIC versus the original data; and VM-NORDIC versus Standard-NORDIC. The random patterns indicate that VM-NORDIC leads to more noise removal, especially in central areas with high g-factor values, without affecting structural details.

*Magnitude images*

Figure 3A shows a reconstructed example slice before denoising and after denoising with Standard-NORDIC and VM-NORDIC, respectively. Both denoising methods lead to improvements upon visual inspection. Particularly, VM-NORDIC reduces noise more effectively in highly noisy central regions (see Figure 4B). The zoomed insets show a more detailed view of the denoising performance of VM-NORDIC: a more drastic noise attenuation than Standard-NORDIC without degrading structural detail.

*Subtraction images*

Figure 3B reports the subtraction images for the same representative slice in Figure 3A of the Standard-NORDIC versus the original image and VM-NORDIC image versus the original image. The two images only show random noise with higher intensities in central regions with high g-factors (Figure 4). In particular, as also visible in the zoomed views, VM-NORDIC leads to a higher degree of noise removal globally and in central areas. The subtraction of VM-NORDIC from the Standard NORDIC image shows that VM-NORDIC removes random noise with an emphasis on noisier central areas.

*tSNR*

Table 1 shows the mean tSNR scores and percentage increases for VM-NORDIC compared to the original data and Standard-NORDIC data for different resolutions and SENSE factors. On average, VM-NORDIC leads to tSNR levels ~9-90% larger than Standard-NORDIC and ~23-250% larger than the original. The highest noise removal occurs for the data with the largest voxel size ($0.7 \times 0.7 \times 0.8 mm^3$, datasets A and B), especially for the dataset with the lower initial tSNR due to the higher SENSE factor (dataset B). Figure 4A shows tSNR maps for two representative slices of dataset B before and after denoising with Standard NORDIC and VM-NORDIC. VM-NORDIC leads to a globally higher tSNR compared to Standard-NORDIC, with a higher increase in central areas coinciding with high g-factor values (Figure 4B). Figure 4B shows the normalized tSNR difference map between Standard NORDIC and VM-NORDIC, the g-factor map estimated in NORDIC via MPPCA, the g-factor map as measured by the scanner and the noise map obtained by acquiring the MR signal without RF excitation of the sample.

*Smoothness estimates*

On average, VM-NORDIC increased spatial smoothness by $4.8 \pm 3.9\%$ (std) compared to the original smoothness estimate. Standard-NORDIC increased spatial smoothness on average by $24 \pm 11.6\%$ (std). The smoothness patterns were similar across all datasets. Figure 5 reports the spatial smoothness estimates for dataset B before and after data denoising with Standard-NORDIC and VM-NORDIC. Here, Standard-NORDIC leads to a substantial smoothness increase, while VM-NORDIC manages to keep smoothness at a level similar to the original.

| VOXEL SIZE | SENSE FACTOR | TSNR | | | TSNR INCREASE (%) | |
|---|---|---|---|---|---|---|
| | | original | Standard NORDIC | VM-NORDIC | original vs. VM-NORDIC | Standard NORDIC vs. VM-NORDIC |
| 0.5×0.5×0.5mm$^3$ (C) | 3 | 7 | 8 | 8.6 | 23 % | 9 % |
| 0.45×0.45×0.45mm$^3$ (F) | 3 | 7.3 | 8.6 | 9.7 | 33 % | 13 % |
| 0.5×0.5×0.8mm$^3$ (D) | 3 | 8 | 10 | 12.6 | 57.5 % | 27.4 % |
| 0.7×0.7×0.7mm$^3$ (E) | 4 | 5.4 | 7.4 | 9.3 | 72.2 % | 25.5 % |
| 0.7×0.7×0.8mm$^3$ (A) | 3 | 14 | 21.3 | 32.2 | 130 % | 51.3 % |
| 0.7×0.7×0.8mm$^3$ (B) | 4 | 10 | 18.4 | 35 | 250 % | 90.2 % |

**Table 1.** tSNR levels for the original data, after Standard-NORDIC denoising and after VM-NORDIC denoising and the percentage tSNR increase after VM-NORDIC with respect to the original data and Standard-NORDIC processed data at different spatial resolutions and SENSE factors.

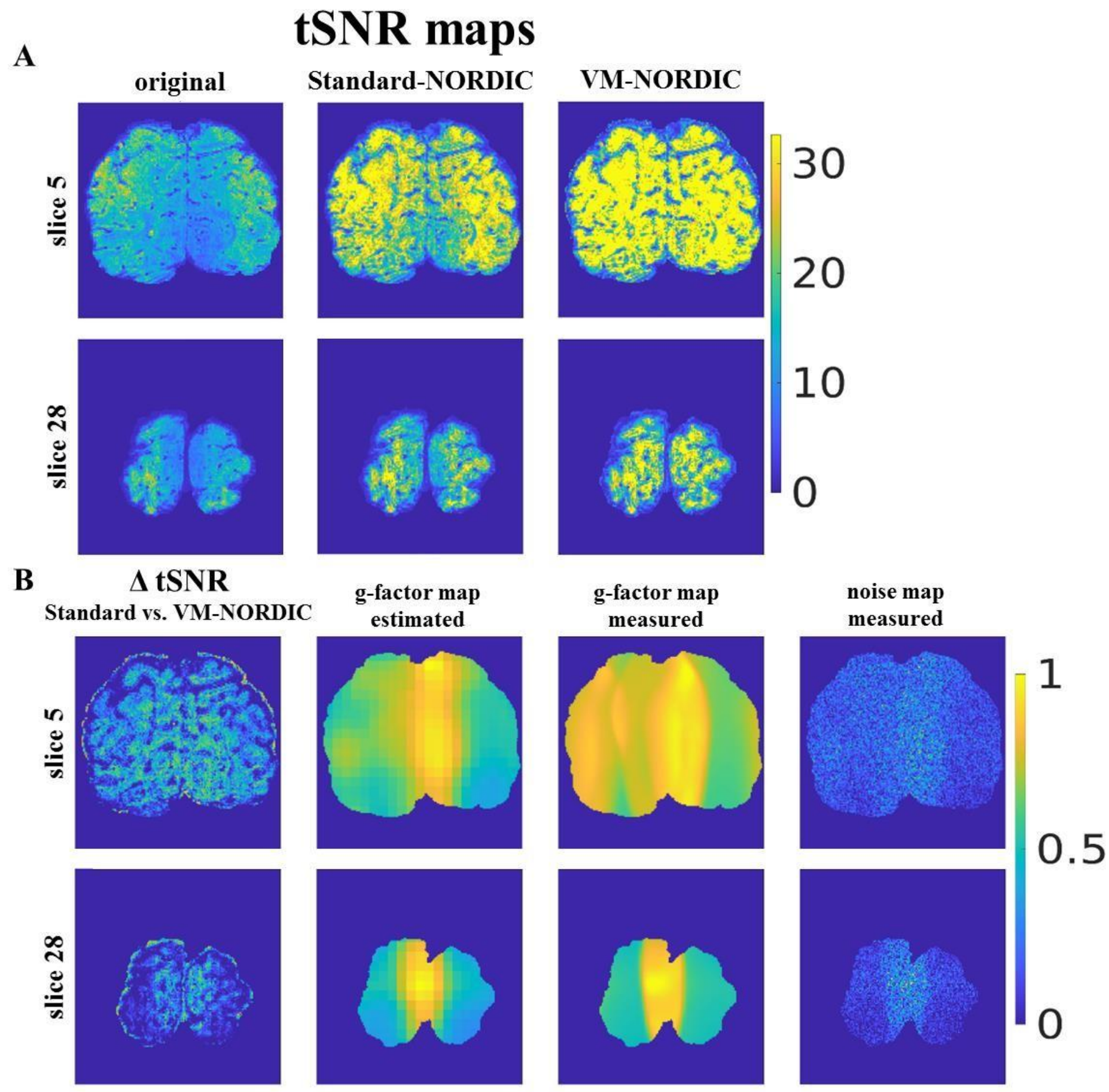

**Figure 4. A)** tSNR maps for two representative slices of dataset B for the original data, Standard-NORDIC denoised data and VM-NORDIC denoised data. **B)** From left to right, the normalized tSNR difference map between Standard-NORDIC and VM-NORDIC; g-factor map estimated in NORDIC via MPPCA; g-factor map measured by the scanner; measured noise map of dataset B. VM-NORDIC increases tSNR levels especially in central areas of the brain with higher g-factor noise amplification.

*Patch size fine-tuning*

Supplementary figures 1 and 2 show how different patch sizes affect the denoising performance in terms of mean tSNR and smoothness scores for the two methods. The “tSNR vs patch size” plots (Suppl. Figure 1) show steep curves with clear maxima for VM-NORDIC within the range of 20-240 time series per patch, with small fluctuations due to the length of the time series (Suppl. Figure 3). Suppl. Figure 2 shows that VM-NORDIC keeps spatial smoothness estimates similar to or slightly higher than the original one but always lower than after Standard-NORDIC. Suppl. Figure 3 and 4 show the tSNR vs. patch size and the smoothness vs. patch size plots for datasets with different time series lengths. All plots in Suppl. Figure 3 show a slight shift of the maxima towards smaller patch sizes with decreasing time series length. Further, shorter datasets have an initially higher tSNR and show a larger tSNR gain following VM-NORDIC. Suppl. Figure 4 shows consistent but non-significant smoothness fluctuations with increasing time series lengths.

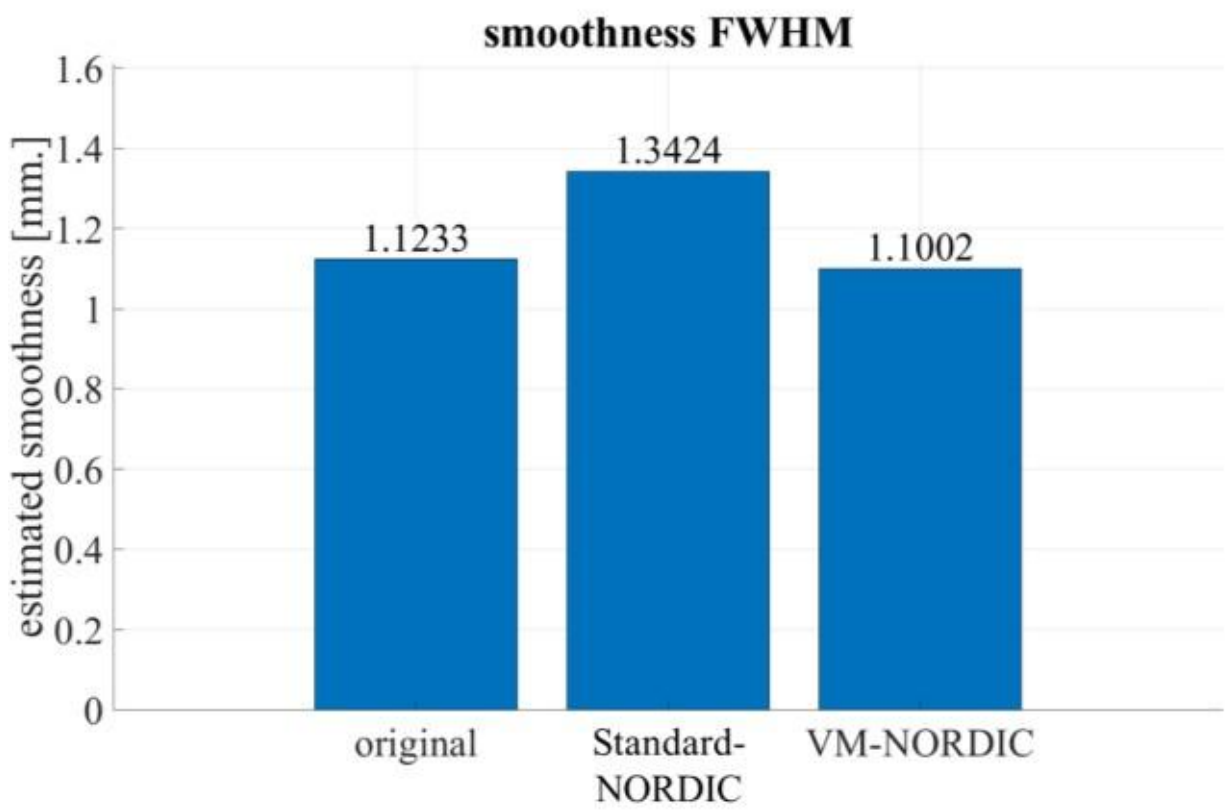

**Figure 5.** Spatial smoothness estimates (using 3dFWHMx AFNI) for the representative dataset B. On average, the proposed VM-NORDIC barely affects the original spatial smoothness of the dataset (~5%), indicating improved preservation of structural detail compared to Standard NORDIC. Specifically, for this dataset, VM-NORDIC results in the spatial smoothness estimate being lower than for the original data. This result follows from the particularly small patch size used on this dataset, which counteracts signal leakage-induced smoothness in the original data.

*tSNR/smoothness*

Suppl. Figure 5 displays the plots of the normalised tSNR/smoothness ratio scores vs. patch size. Here, the maxima in the curves indicate the patch sizes with which the methods better remove random noise without excessively smoothing the data. The maxima fall within the same range reported previously (20-240). Also, VM-NORDIC always surpasses Standard-NORDIC in terms of this metric.

## DISCUSSION

We introduced VM-NORDIC, an extension to the NORDIC PCA algorithm that further decreases noise in BOLD fMRI data. VM-NORDIC achieves this result via non-local patch formation using voxel similarity matching to boost the low-rank properties of the patches, which facilitates denoising performance using
SVT. The Standard-NORDIC and VM-NORDIC patches before and after denoising in Figure 2 show how the higher homogeneity of the VM-NORDIC patch allows emphasizing the signal fluctuations that the time series have in common, visible as fading vertical stripes. Better-defined signal fluctuations enable the uncovering of the underlying temporal structure of the brain activity, which otherwise would be too contaminated by noise (Figure 2B). Overall, VM-NORDIC reaches a higher degree of thermal noise removal, as indicated by substantial increases in tSNR (up to 2-fold the original tSNR), while better preserving spatial specificity and structural detail, which are key advantages in submillimeter resolution BOLD fMRI studies. Altogether, these results indicate that non-local patching promotes a superior low-rank structure of the data, allowing the data to be adequately represented with fewer principal components than with local patches in Standard NORDIC. These improvements are in line with the results reported by Zhao et al., where patching using non-local similarities to denoise dMRI and DTI data significantly reduced noise while preserving structural details compared to MPPCA (Zhao et al., 2022). As mentioned by the original NORDIC authors (Vizioli et al., 2021), NORDIC is beneficial and complementary to other pre-processing steps that target different aspects of the dataset, such as motion correction and physiological noise removal.

The tSNR scores in Table 1 report that the effectiveness of VM-NORDIC increases with increasing voxel size. We argue that these results derive from the higher signal strength at greater voxel sizes, leading to larger and more pronounced principal components containing signal, which benefits SVT. Notably, among the two datasets with the largest and equal voxel size (datasets A and B), the one with the initially higher noise level (dataset B, due to the higher SENSE factor) shows the largest tSNR gain. This result implies that VM-NORDIC targets thermal noise only while leaving the target signal intact. However, the higher noise level of dataset B due to the greater SENSE factor makes its tSNR lower than in dataset A. Consequently, as visible in Table 1, the two datasets reach similar tSNR levels upon VM-NORDIC, implying that denoising removed a larger amount of thermal noise from the noisier dataset B.

That VM-NORDIC effectively removes more noise while preserving the structural integrity of the images is visually perceivable in the reconstructed images (Figure 3A). Moreover, the difference between VM-NORDIC and the original image shows only noise without notable edge effects (Figure 3B). The difference between VM-NORDIC and Standard-NORDIC also shows mostly noise, indicating that non-local patching in VM-NORDIC removes additional noise than Standard-NORDIC. Moreover, the same image also exhibits spatial patterns in correspondence with high g-factor values (Figure 4B). These results suggest that the higher mean tSNR after VM-NORDIC mainly derives from stronger denoising of central areas. Nonetheless, peripheral areas also exhibit a less pronounced yet tangible tSNR increase after VM-NORDIC, indicating a globally stronger noise removal (see Figure 4).

These converging findings point out that in local patches from highly noisy regions, noise irreversibly propagates through all signal components, leading to less effective SVT compared to areas where noise is less dominant. This scenario is never obtainable with VM- NORDIC since non-local patches inherently have well-pronounced signal components by construction. Always having well- pronounced signal components allows VM-NORDIC to denoise all voxels across the dataset with comparable efficiency, as long as enough similar voxels are present to generate sufficient redundancy. Therefore, the outcome is, on average, a high and consistent tSNR level across the whole dataset independently of the local noise levels.

VM-NORDIC also induces lower levels of spatial smoothing than Standard-NORDIC (Figure 5, Suppl. Figure 2). As mentioned, Standard-NORDIC applies an LLR method that inevitably increases spatial smoothing by increasing the similarity between adjacent voxels (see Methods and Materials, section "*Standard-NORDIC LLR model*"). Importantly, we observe that VM-NORDIC minimally smooths the dataset by denoising groups of voxels from different locations across the dataset, increasing the original smoothness by less than 5% on average (about 20% of that induced by Standard NORDIC). Minimal contamination of the spatial integrity is crucial for sub-millimeter BOLD fMRI applications. The steep tSNR curves as a function of patch size in Suppl. Figure 1 suggest that the performance of VM-NORDIC is more sensitive to the patch size compared to Standard-NORDIC. Notably, for VM-NORDIC, the sharp maxima of mean tSNR as a function of patch size reveal the existence of an optimal patch size per dataset. We argue that these results derive from the notion that in Standard-NORDIC, adding or subtracting a few voxels from an inhomogeneous local patch does not strongly modify its low-rank representation and thus the effectiveness of denoising (Suppl. Figure 1). Conversely, in VM-NORDIC, having larger patches can lower the overall similarity ranking, which potentially reduces the patch homogeneity and hampers efficient SVT. However, larger patches can also improve denoising by boosting signal redundancy if the additional voxels are sufficiently similar to the reference one. Oppositely, small patches may be highly homogeneous but not exhibit enough redundancy for a reliable SVT. These concepts suggest that the optimal patch size is a data-driven trade-off between the degree of similarity of the voxels and the level of signal redundancy. Hence, a data-specific fine-tuning process is necessary to find the optimal patch size (see Methods and Materials section "*Patch size optimization*"). The figure also shows that VM-NORDIC produces higher tSNR scores than Standard-NORDIC at suboptimal patch sizes, meaning that the patch size alone is not the sole promoter of the improvements. Suppl. Figure 3 illustrates that the optimal patch size per dataset slightly fluctuates or decreases with decreasing time points, and that data with fewer time points reach a higher tSNR increase thanks to the higher initial tSNR.

Smoothness scores per different patch sizes indicate that, overall, the degree of smoothing upon VM-NORDIC remains lower than for Standard-NORDIC independently of the size of the time series (Suppl. Figure 2 and 3). In particular, in VM-NORDIC, small patch sizes generally preserve spatial detail more effectively than large ones. We suggest that the reduced number of time series allowed in small patches limits the chance of denoising the same time series too many times, which could potentially remove the characteristic underlying signal. Further, signal leakage due to, for instance, susceptibility artefacts can slightly increase the smoothness of the original data by spreading the magnetic signals from one brain region into neighbouring voxels, leading to a decrease in spatial resolution and blurring[39]. Consequently, particularly small patches avoid further signal spreading or, in a few cases, even counteract signal leakage effects by decreasing the artificial similarities between adjacent voxels, leading to comparable or lower smoothness estimates than in the original data (e.g. Figure 5, Suppl. Figure 2, second panel).

Since the goal of a denoising method is maximizing the tSNR while minimizing spatial smoothing, combining tSNR and smoothness scores into a single metric allows for a direct and global assessment of the denoising performance. The normalized tSNR/smoothness ratio vs. patch size plots for different time series in Suppl. Figure 5 shows that VM-NORDIC reaches higher tSNR/smoothness ratios than Standard NORDIC, thus a more convenient trade-off between noise removal and induced smoothness. Further, also these curves emphasize a range of dataset-specific optimal patch sizes (20-240) that guarantee a higher degree of noise removal while preserving structural information.

Ideally, a reference voxel should be compared to all brain voxels across the entire dataset to maximize the chance of finding high similarities and grouping voxels from different slices. However, computing the similarity scores of the whole dataset at once is computationally expensive. A solution was to process less data per cycle by dividing the dataset into equally large chunks of temporal slices and denoise each chunk individually (for more details, see the Methods and Materials section "*Chunk Size*"). As long as the chunks contain enough data and slices, there will still be a high chance of finding highly similar non-local voxels. After selecting the appropriate chunk size based on the performance of our machines, we saw that chunking significantly speeds up the processing without interfering with the quality of denoising.

Finally, a recent study reported that Standard NORDIC and MPPCA occasionally generate artificial functional activation depending on the selected patch size[11]. In particular, they found that larger patch sizes provide higher sensitivity to BOLD responses with both MPPCA and Standard-NORDIC, but with significant activation "spreading" and increasing false-positive rate due to the local bleeding of active signal components[11]. They concluded that for both methods, the optimal patch size for each experiment depends on data tSNR and functional CNR. Similarly, our findings also show an increase in the induced spatial smoothness (hence signal spreading) with increasing patch sizes for VM-NORDIC (Supp. Figure 2). However, as discussed, the patch size optimization process in VM-NORDIC already relies on the data tSNR and usually estimates patch sizes smaller than those used in Standard NORDIC. Hence, we suggest that the lower spatial smoothness in VM-NORDIC with small patch sizes can limit the activation spreading by keeping unaltered or even decreasing the artificial spatiotemporal similarities between adjacent voxels (Risk et al., 2018). Yet, a more dedicated investigation is necessary to understand if a similar effect occurs during VM-NORDIC even after the patch size optimization.

Among the limitations of VM-NORDIC, there is the inevitable increase in the processing time (on average ~1.8 times longer than with NORDIC standard, but still within ~5-10 minutes on our machines depending on the size of the data) due to the additional steps of voxel matching and patch size fine-tuning. Further, the present study did not include analyses of functional activity such as t-statistic activation maps. Since the end goal of VM-NORDIC is to improve the reliability and further push sub-millimeter BOLD fMRI studies, these statistical analyses are of paramount importance. However, we suggest that because the voxels in a VM-NORDIC patch are already highly correlated owing to their similar time courses, increasing their similarity may help reveal unspotted activation.

## CONCLUSION

A large part of the neuroscientific community is devoted to studying mesoscopic cortical organizations like cortical columns and layers in the visual cortex (Dumoulin et al., 2018). So far, most of the detailed studies were possible only by invasively mapping animal cortices. Expanding the investigation to in vivo human brains requires noninvasive, precise and reliable imaging methods (Bandettini et al., 1992). Specifically, the BRAIN Initiative Working Group reported that the voxel size (e.g. resolution) necessary to resolve these small-scale structures spans from ~0.46 mm to ~0.55 mm isotropic (Jorgenson et al., 2015). Reducing noise while preserving spatial integrity is essential to enable high-resolution fMRI studies revealing new neuroscientific insights at the mesoscopic scale. Together with the latest advancements in hardware technologies, such as the development of dedicated receive head coils, powerful gradient inserts, and even stronger static magnetic fields (Petridou et al., 2013; Sengupta et al., 2016; Versteeg, Klomp, et al., 2021; Versteeg, van der Velden, et al., 2021) VM-NORDIC will allow the community to further push the boundaries of fMRI resolution (Petridou et al., 2013; Viessmann & Polimeni, 2021). Future steps should investigate and validate the effects of VM-NORDIC on functional mapping and apply it to higher-resolution data acquired with the aforementioned specialised hardware.

## ACKNOWLEDGEMENTS

The authors thank Stanley D.T. Pham for assistance with the 7T MRI system and acquisition software, and Jason van Schoor for technical support. High-resolution fMRI data were provided by the Translational Neuroimaging Group, UMC Utrecht. Portions of this work were presented at ISMRM 2023 (Toronto) and the ISMRM Benelux Chapter 2023 (Brussels). This work was supported by the National Institute of Mental Health of the National Institutes of Health under the Award Number R01MH111417 and the Rudolf Magnus Fellowship (J Siero) of the University Medical Center Utrecht, the Netherlands.

## SUPPLEMENTARY MATERIALS

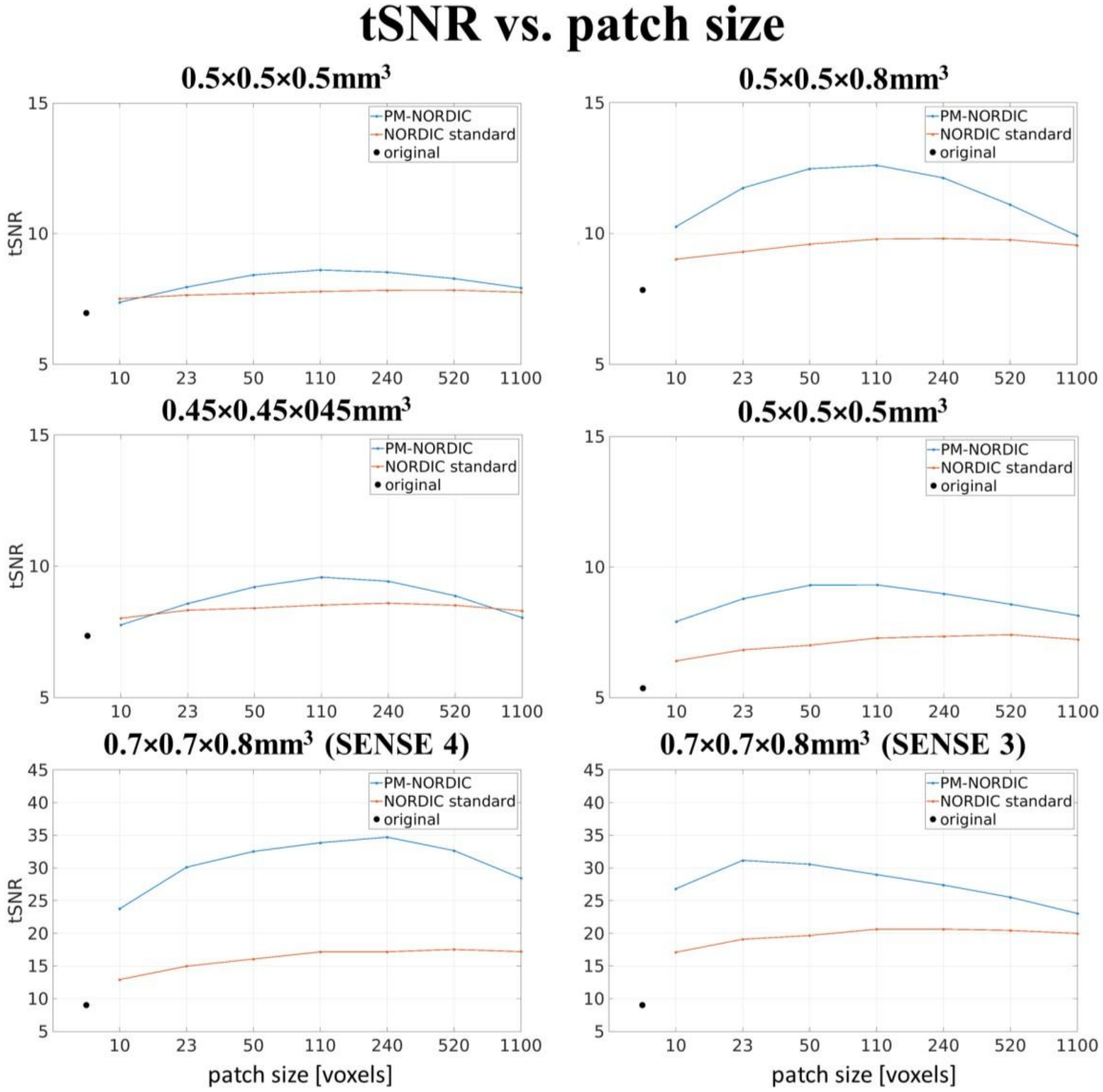

**Supplementary Figure 1.** Mean tSNR as a function of patch size for Standard-NORDIC and VM-NORDIC. The denoising performance of VM-NORDIC is more sensitive to the chosen patch size. Also, The curves for VM-NORDIC show clear maxima representing the optimal patch sizes.

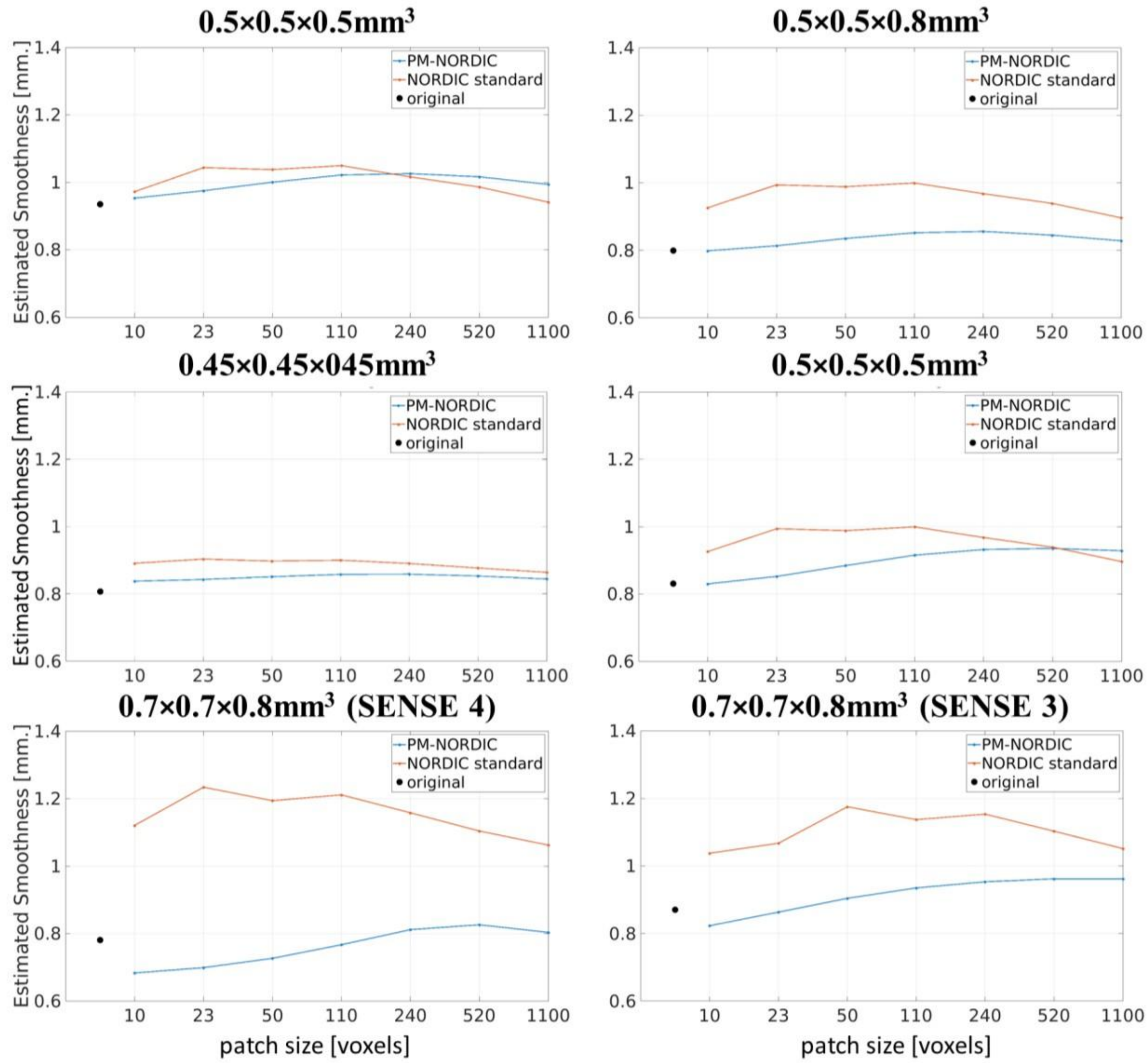

**Supplementary Figure 2.** Global smoothness estimates (FWHM) as a function of patch size for Standard-NORDIC and VM-NORDIC. VM-NORDIC does not significantly increase spatial smoothing and, for certain patch sizes, even leads to smoothness estimates lower than the originals by counteracting smoothing signal leakage.

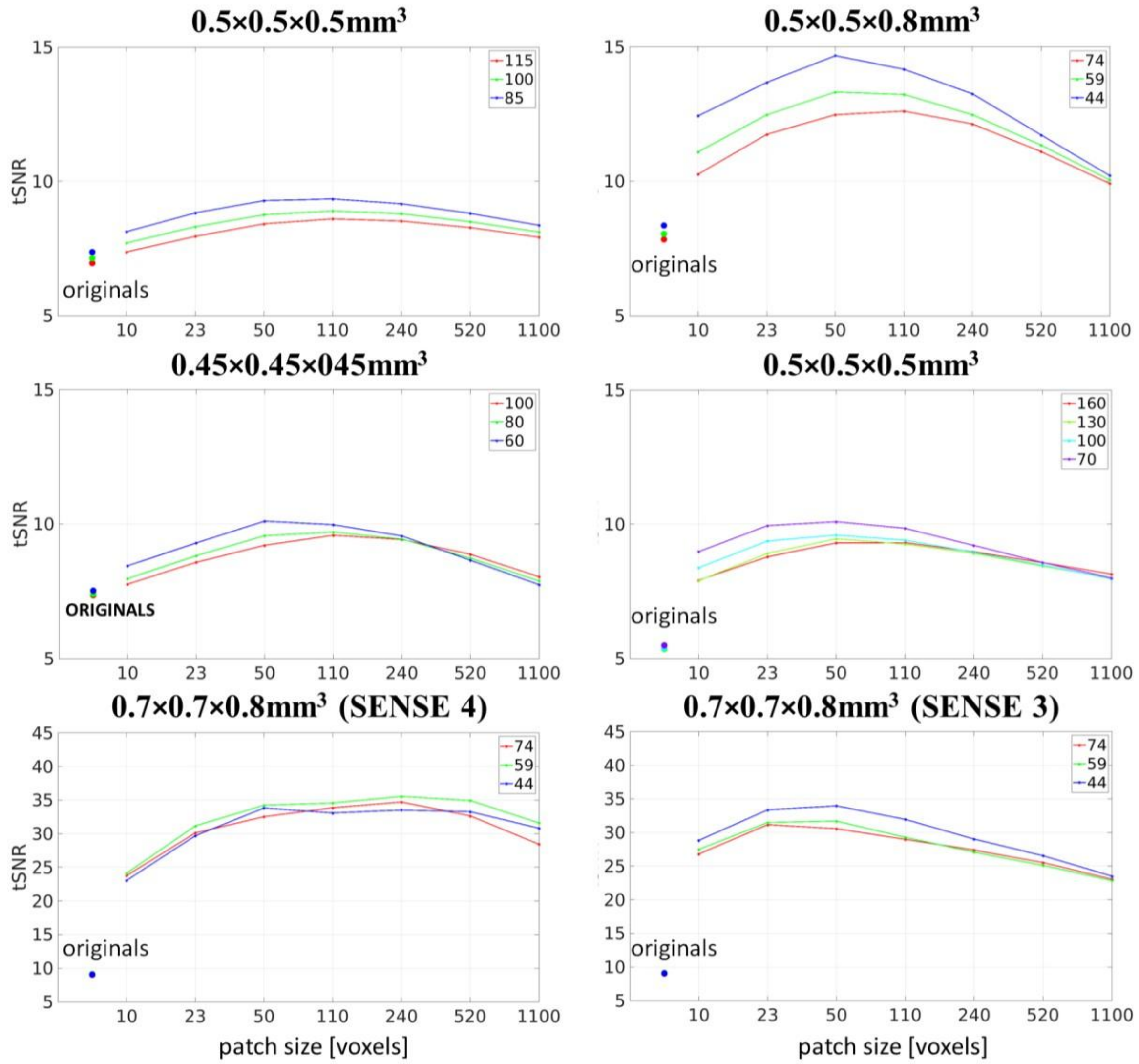

**Supplementary Figure 3.** Mean tSNR as a function of patch size for VM-NORDIC denoised datasets with different numbers of time points. The optimal patch size slightly increases with increasing time points. Nonetheless, the shift is not significant as long as the number of time points is within a regular range. Additionally, shorter datasets show the largest tSNR gains.

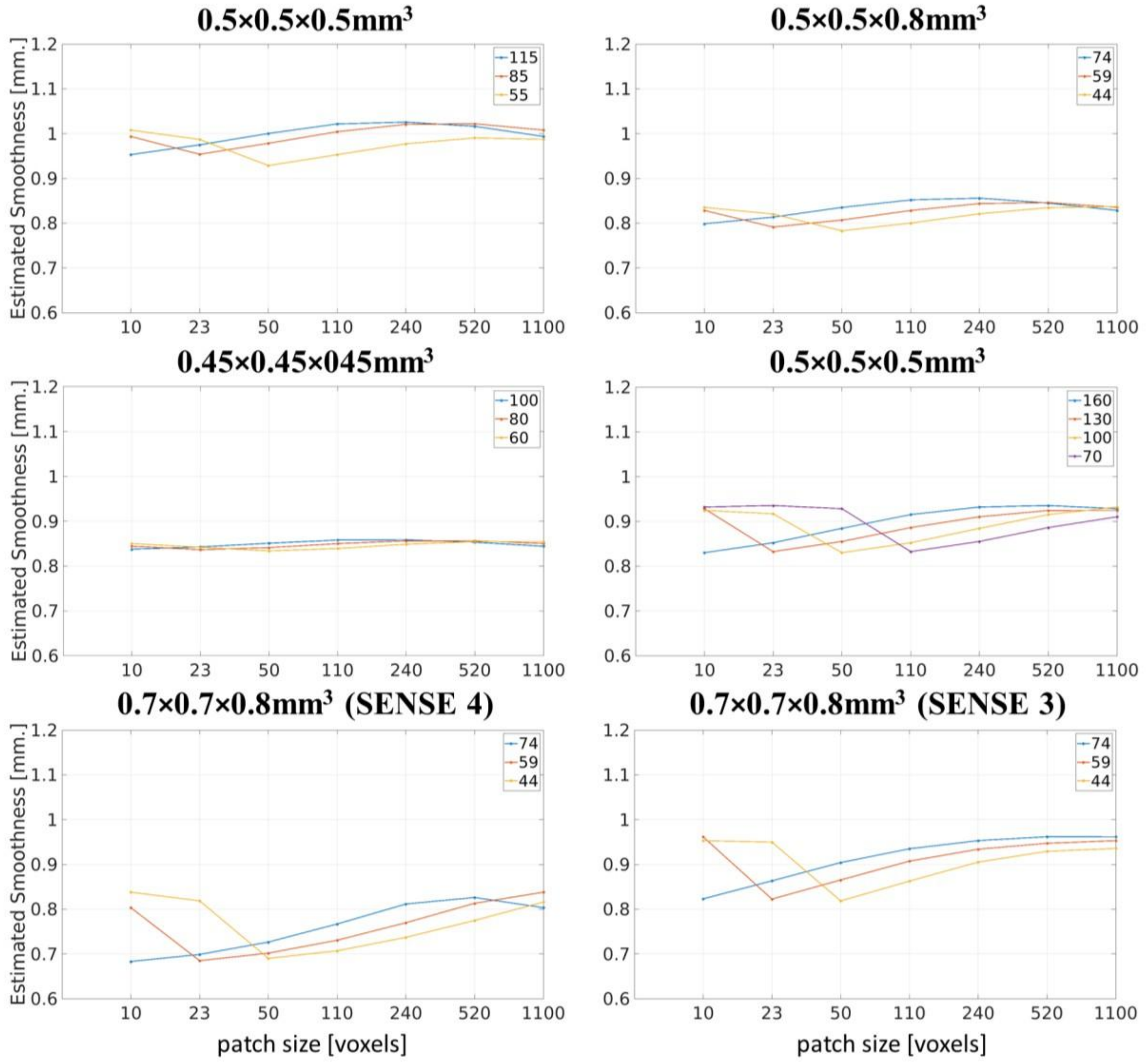

**Supplementary Figure 4.** Global smoothness estimates (FWHM) as a function of patch size for VM-NORDIC denoised datasets with different numbers of time points. The degree of spatial blurring exhibits small but irrelevant fluctuations as the number of time points of the dataset changes.

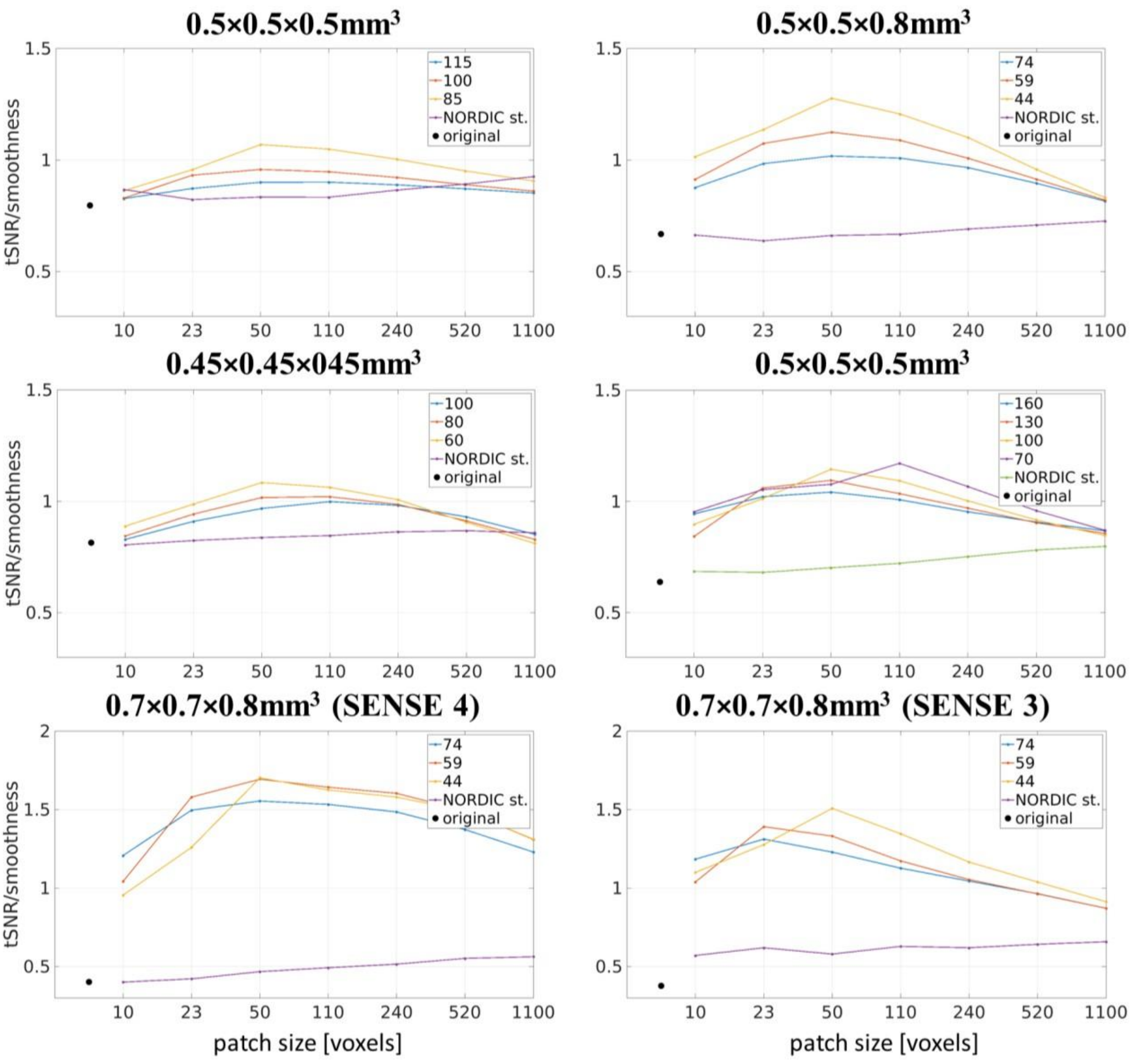

**Supplementary Figure 5.** Normalized tSNR over normalized global smoothness estimate (FWHM) ratio as a function of patch size for VM-NORDIC denoised datasets with different number of time points. This metric shows which patch size guarantees the best denoising performance in terms of the trade-off between nosie removal and induced spatial smoothness. The curves indicate that VM-NORDIC has a clear advantage over Standard-NORDIC, and that the optimal patch sizes range from 20 up to 240 timeseries per patch.